\documentclass[epj,twocolumn]{webofc}
\usepackage[varg]{txfonts}
\usepackage{subfigure}
\usepackage{hyperref}
\woctitle{TRANSVERSITY 2014}
\begin{document}
\title{Transverse Momentum Effects in Unpolarised SIDIS at COMPASS}
\author{Nour Makke\inst{1,2}\fnsep\thanks{\email{nour.makke@cern.ch}}}
\institute{Universit\`a degli studi di Tieste, Dipartimento di Fisica, Via A. Valerio 2, 34127 Trieste, Italy
\and INFN sezione di Trieste, Padriciano, 99, I - 34149 Trieste, Italy
}

\abstract{Unpolarised semi-inclusive DIS is receiving a growing interest as a powerful tool to access poorly known universal functions that play a key role in many processes, in particular in the study of the spin structure of the nucleon. These functions can be investigated through experimental observables such as hadron multiplicities in the collinear and transverse framework, the dihadron multiplicities, the azimuthal asymmetries and some others. New results on these observables by the COMPASS experiment at CERN will be shown and discussed.}

\maketitle

\section{Introduction}
\label{sec-0}

Over the last few decades, a major effort has been dedicated, both on experimental and theoretical sides, to the study of the nucleon spin structure by determining the parton helicity and transversity distributions, considering that the internal 2$-$D partonic structure of the nucleon is precisely known except the strange quark density, which still carries a large uncertainty. The transverse momentum structure of the nucleon, however, was not considered with the same level of priority although it is poorly known. Recently, a considerable effort is being dedicated to achieve a good level of understanding of the internal 3$-$D partonic structure of the nucleon. Thus a growing interest in the determination of the Transverse Momentum Dependent (polarised and unpolarised) Parton Distribution Functions (PDFs) and Fragmentation Functions (FFs) is taking place. 

The most viable and promising process to assess/determine $k_{\perp}$ and $p_{\perp}$ and their kinematical dependencies as well as their dependencies upon quark charges and flavours is the Semi-Inclusive Deep-Inelastic Scattering reaction. In the framework of perturbative QCD (pQCD) at leading twist, SIDIS is viewed as the hard scattering of a lepton off a quark or antiquark, which subsequently hadronizes into final state hadrons. The hard-scattering cross section is calculable in pQCD while the PDFs and FFs are non-perturbative quantities. In the collinear framework, i.e. integrated over transverse momentum, they are believed to be universal, i.e, process-independent. In this case, PDFs are known with a high precision while FFs are known for pions and poorly known for kaons. With the inclusion of the kinematic dependence on the hadron transverse momentum, beyond the standard collinear factorisation is reached and the transverse momentum dependence of PDFs and FFs is assessed. While the TMD-PDFs parametrize the flavour transverse momentum structure of the nucleon and encode details about the motion of quarks inside a nucleon due to their intrinsic transverse momentum (referred to as $k_{\perp}$), the TMD-FFs encode details about producing a final-state hadron with a component of hadron momentum transverse with respect to the momentum transfer (referred to as $p_{h\perp}$) different from zero.

In addition to the many advantages that SIDIS offers as flavour separation, its relevance for spin physics and its accessibility to large hadron energy fraction, it covers a wide range in the energy scale (accessing smaller $Q^2$) allowing to improve the scale coverage in the $Q^2$ evolution, essential to determine gluon contribution. SIDIS measurements profit from the different probabilities to produce a hadron of certain specie from a parton of a given flavour and have been successfully performed at fixed-target experiments for $k_{\perp}$ \& $p_{\perp}$ integrated observables. 

In this paper, we present recent measurements of experimental observables, which encode details of the transverse momentum effects in SIDIS, using data samples collected by the COMPASS collaboration at CERN. Among these observables are the $p_{T}^{2}$-dependent multiplicities and the azimuthal asymmetries of unidentified hadrons, both measured in unpolarised semi-inclusive DIS. In sec. \ref{sec-1}, the experimental setup is briefly described, the hadrons multiplicities (analysis and results) are presented in sec. \ref{sec-2} and the azimuthal asymmetries are presented in sec. \ref{sec-3}.
 
\section{The Experiment}
\label{sec-1}
 
COMPASS is a fixed target experiment, which uses the muon beam delivered by the CERN Super Proton Synchrotron (SPS). The momentum of the beam is determined at the end of the beam line with a precision of 0.3\% and its trajectory is measured before the target in a set of silicon and scintillating fibre detectors. The target consists of three cells, located among the beam one after the other, filled with $^6$LiD immersed in a liquid $^3$He/$^4$He mixture. The two-stage COMPASS spectrometer is designed to reconstruct scattered muons and produced hadrons in a wide range of angle and momentum. Particle tracking is performed by a variety of tracking detectors before and after the two spectrometer magnets. The direction of the reconstructed tracks at the interaction point is determined with a precision of 0.2 mead while the momentum resolution is 1.2\% in the first spectrometer stage and 0.5\% in the second. The trigger is formed by hodoscope systems supplemented by hadron calorimeters. 

\section{Hadron Multiplicities}
\label{sec-2}
The hadron multiplicity is the most suitable observable allowing to access TMD parton densities and fragmentation functions simultaneously. In the TMD factorisation scheme, the spin-averaged differential cross section of the SIDIS process for the production of a hadron $h$ with transverse momentum $p_T$ and fractional energy $z$ at leading order in $\alpha_{s}$ and ($k_{\perp}/Q$) reads:

\begin{equation}
\frac{d\sigma^{h}}{dx_{B}dQ^{2}dzdp_{T}^{2}} = \frac{2\pi^2\alpha^2}{(x_{B}s)^2} \frac{[1 + (1-y)^2]}{y^2} F_{UU}
\end{equation}

\begin{equation}
\begin{gathered}
F_{UU} = \sum_{q} e_{q}^{2} \int d^2k_{\perp} d^2p_{\perp} f_{q/p} (x_{B},k_{\perp}) \\
\hspace{1.6cm}D_{h/q}(z,p_{\perp})\delta^{(2)}( p_{T} - zk_{\perp} - p_{\perp})
\end{gathered}
\end{equation}

The TMD-PDFs and TMD-FFs can be extracted using experimental measurement of hadron multiplicities and assuming dedicated functional forms for the TMDs and that, in the $\gamma^*-p$ c.m. frame, the measured transverse momentum of a hadron ($p_{T}$) is generated by the quark intrinsic transverse momentum  ($k_{\perp}$) and of the final hadron with respect to the fragmenting quark ($p_{\perp}$).

\begin{figure*}[htdp]
\centering
\subfigure{\includegraphics[height=8.7cm,width=.327\textwidth]{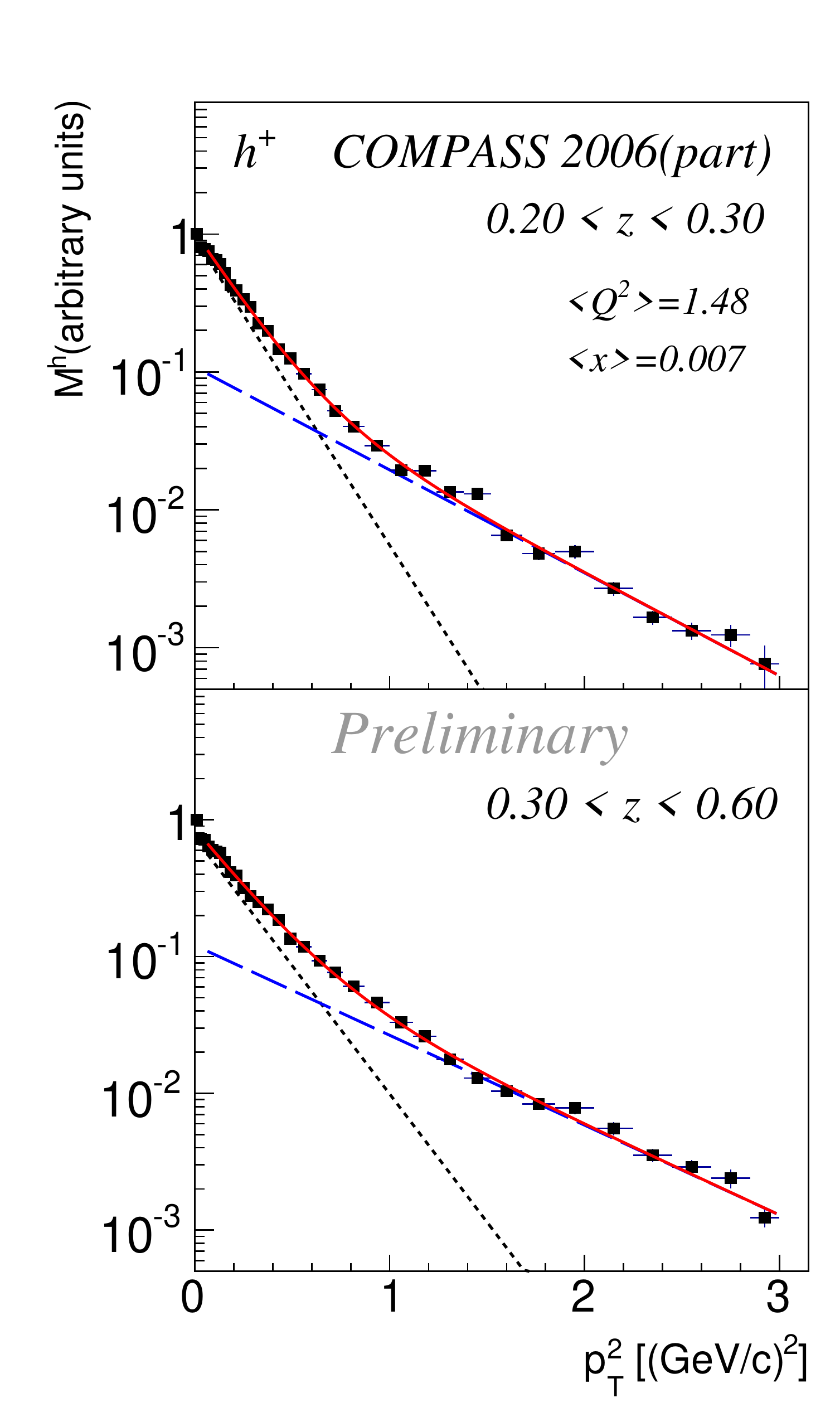}}
\subfigure{\includegraphics[height=8.7cm,width=.327\textwidth]{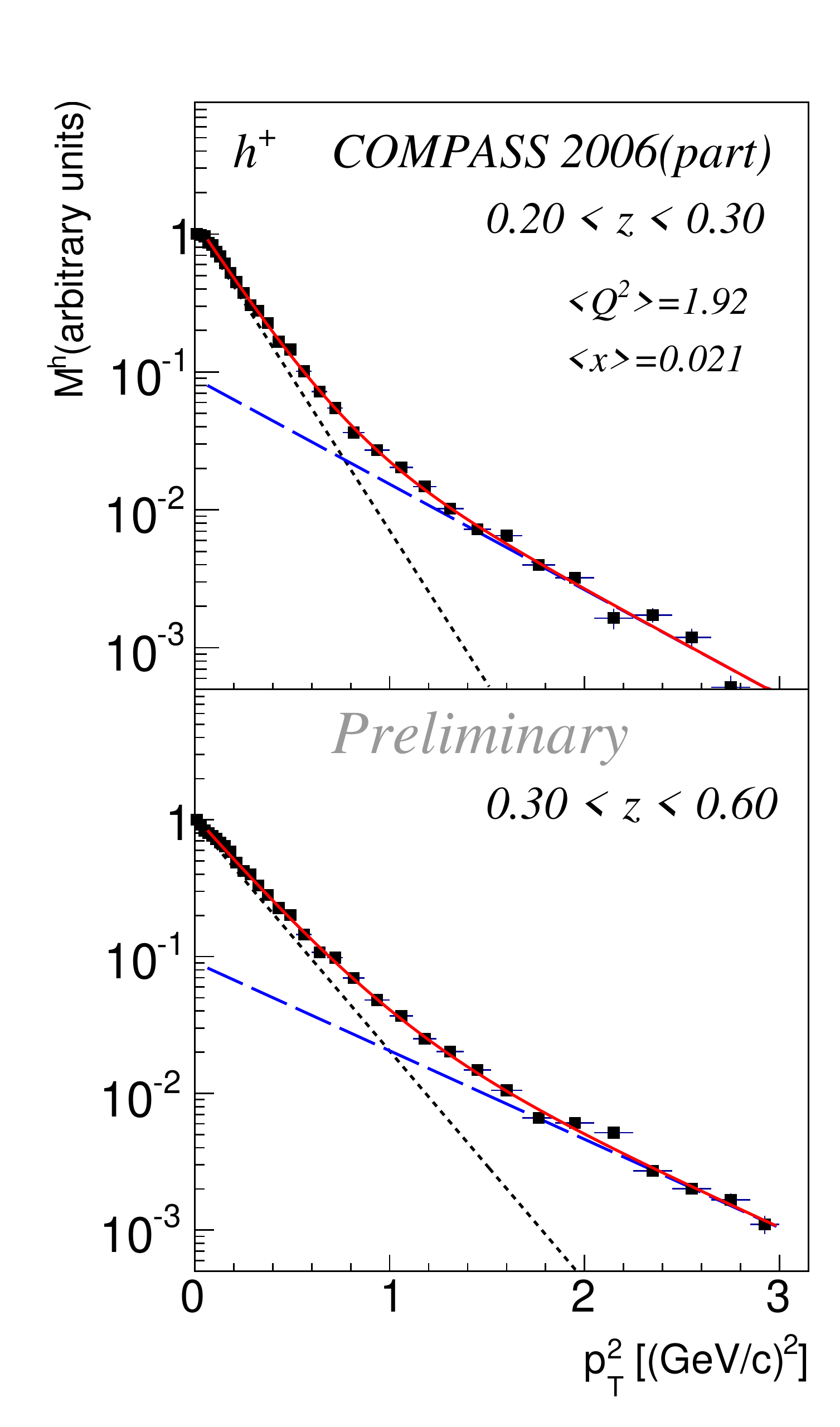}}
\subfigure{\includegraphics[height=8.7cm,width=.327\textwidth]{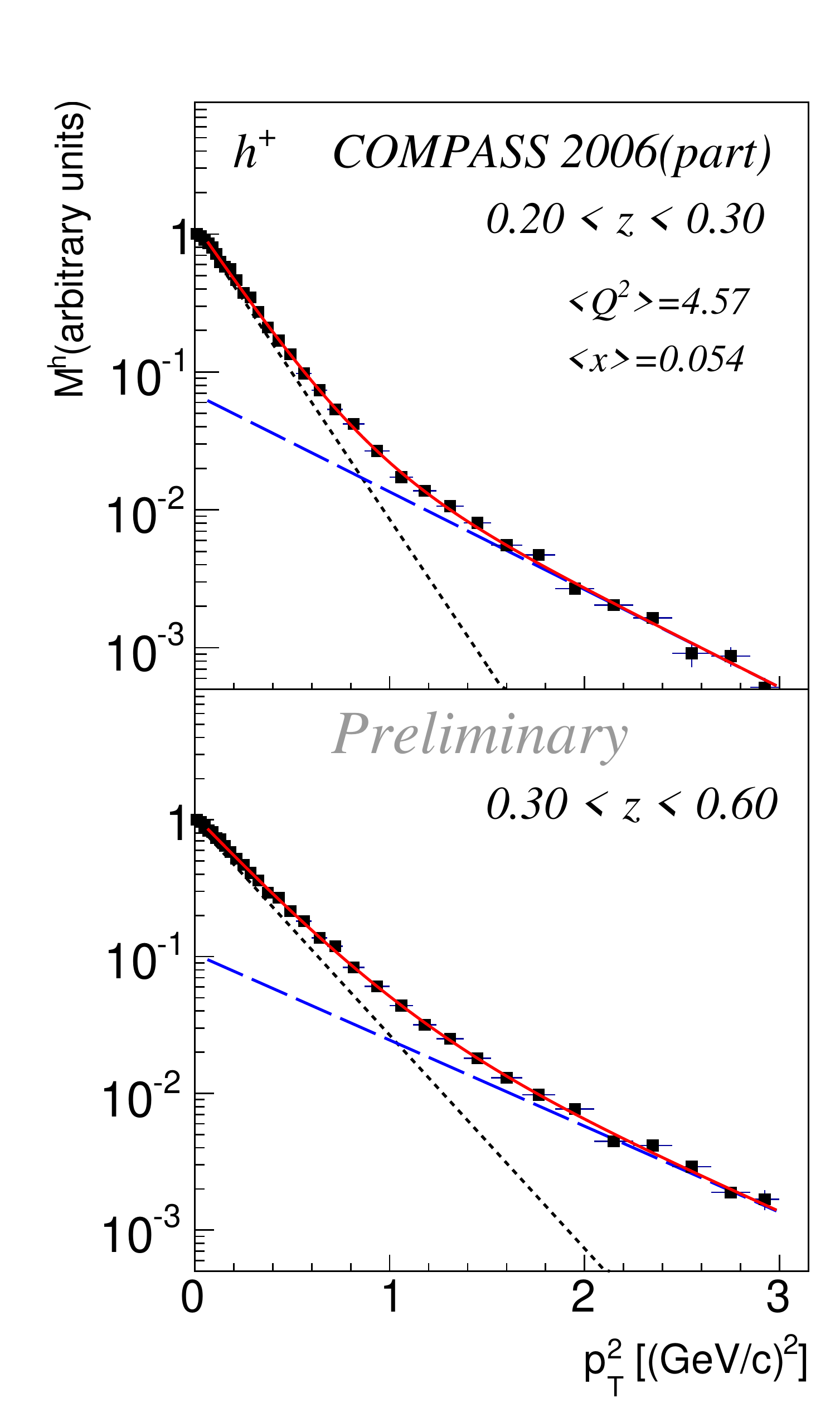}}
\caption{$h^{+}$ $p_T^2$-dependent multiplicities (scaled to arbitrary units) with 1-exponential (dotted line) and 2-exponentials (dashed line) fits. $x$ refers to the kinematic variable $x_B$. Only statistical errors are shown.}
\label{MulFitp}
\end{figure*}
\begin{figure*}[htdp]
\vspace{-0.9cm}
\centering
\subfigure{\label{}\includegraphics[height=8.7cm,width=.32\textwidth]{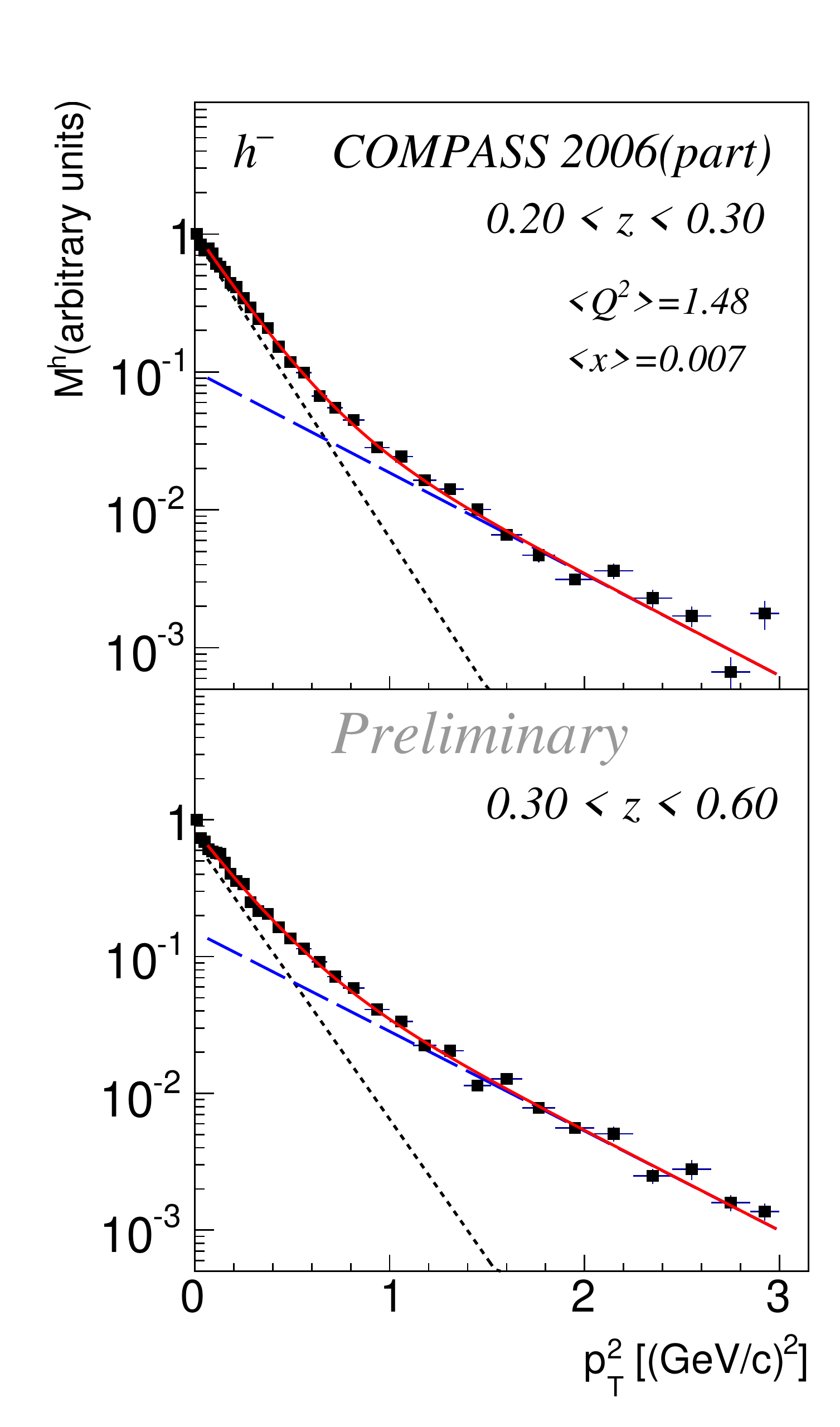}}
\subfigure{\label{}\includegraphics[height=8.7cm,width=.32\textwidth]{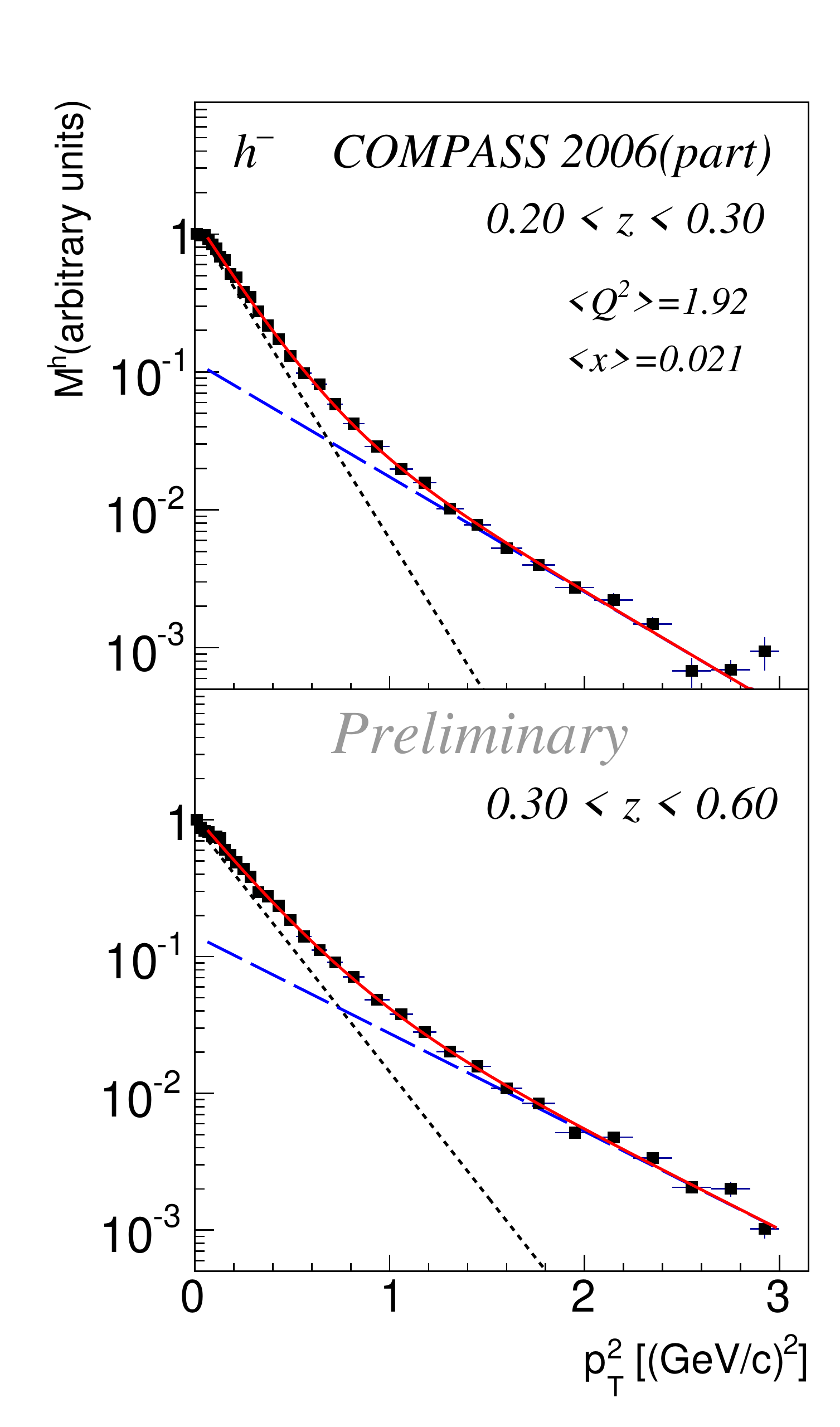}}
\subfigure{\label{}\includegraphics[height=8.7cm,width=.32\textwidth]{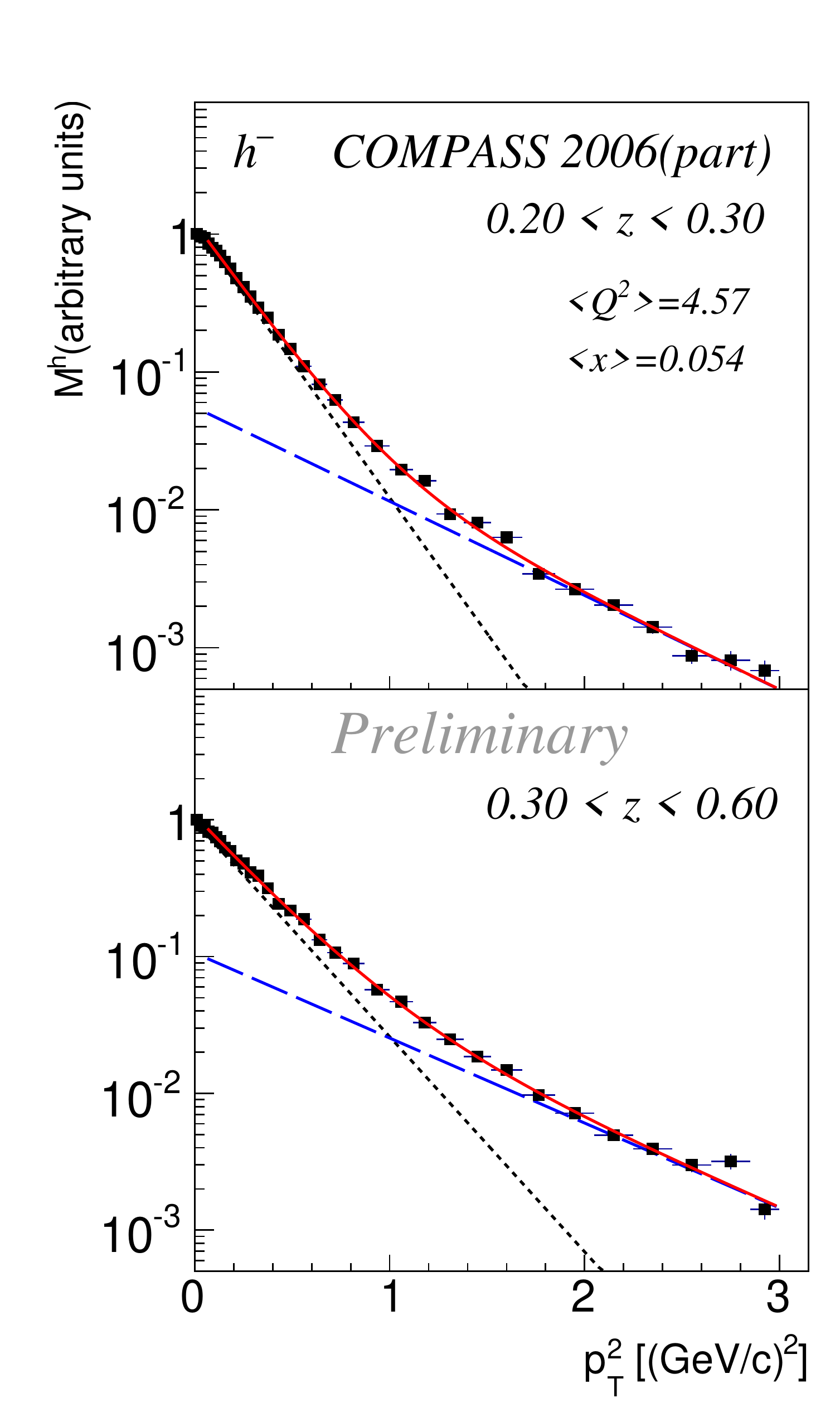}}
\caption{$h^{-}$ $p_T^2$-dependent distributions with 1-exponential (dotted line) and 2-exponentials (dashed line) fits. $x$ refers to the kinematic variable $x_B$. Only statistical errors are shown.}
\label{MulFitm}
\end{figure*}

Previous measurements of $p_{T}$-dependent hadron multiplicities have been performed at fixed target experiments at HERMES \cite{Airapetian:2012ki} and COMPASS \cite{Adolph:2013stb}. The latter used  2004 data collected by scattering $160$ (GeV/c) muon beam off a deuteron ($^6$LiD) target, and was performed in a limited kinematic regime ($0.004<x<0.12$, $Q^2 < 10$ (GeV/c)$^2$, $p_T^2 < 1.3$ (GeV/c)$^2$). due to the limited angular acceptance of the experiment. An upgrade of the COMPASS spectrometer was performed in 2005 allowing a larger geometrical/angular acceptance and an improved RICH PID efficiency. Data collected in 2006, covering a wider kinematic regime (larger $x_B$ and $Q^2$) with an extended range in $p_T^2$ up to $3$ (GeV/c)$^2$, are analysed and the results are presented in this paper. 

A hadron multiplicity is defined as the average density number of hadrons produced per Deep-Inelastic Scattering (DIS) reaction. Experimentally (eq.\ref{mul}), it is defined by the yields of hadrons produced in a selected DIS sample to the number of DIS interactions recorded in this sample. It can be extracted in bins of different kinematic variables providing a multivariate experimental data appropriate for phenomenological extraction of TMDs. 

\begin{equation}
M^{h}(x_B,Q^2,z,p_T^2) = \frac{1}{\frac{d^2n^{DIS}(x_B,Q^2)}{dx_BdQ^2}} \frac{d^4n^h(x_B,Q^2,z,p_T^2)}{dx_BdQ^2dzdp_T^2}
\label{mul}
\end{equation}

For data selection, a DIS event is required to have a primary interaction vertex lying in the fiducial target volume and associated to a muon beam track with energy in the range [140 GeV/c, 180 GeV/c] and a scattered muon track. The DIS regime is defined by the photon virtuality threshold $Q^2 > 1$ (GeV/c)$^2$. Furthermore, a cut $W > 5$ (GeV/c) is required to avoid the region of nucleon resonances. The fractional muon energy transferred to the virtual photon is required to be in $[0.1,0.9]$. While the lower cut avoid kinematic regions where the resolution of muon track reconstruction degrades, the upper cut excludes kinematic regions highly affected by radiative effects. Finally the covered $x_B$ range is [0.003,0.7]. The hadron selection is performed on top of the DIS event selection. 

The measured multiplicity is subject to many effects: limited kinematic and geometric acceptance effects, smearing effects, radiative effects and should be corrected for all of them. The correction for geometric acceptance and smearing effects is the highest correction and is evaluated using a Monte Carlo (MC) simulation composed of a physics generator, a GEANT simulation of track generated leptons and hadrons through a model of the apparatus and finally an algorithm for track reconstruction identical to the one used for experimental data. As the MC simulation contains both the generated and reconstructed properties of all recorded tracks, the acceptance correction factor is defined by the ratio of the reconstructed hadron yield to the generated one, both recorded from the selected DIS sample in simultaneous bins of $x_B$ and $Q^2$ , $z$ and $p_T^2$ as shown in eq. \ref{accp}, where $i$($j$) refers to the bin number based on reconstructed (generated) properties of lepton and hadron MC tracks for each variable. 

\vspace{-0.3cm}
\begin{equation}
\epsilon^{h}(x_B,Q^2,z,p_T^2) = \left. \frac{n^h_r(i_{x_B},i_{Q^2},i_z,i_{p_{T}})}{n^h_g(i_{x_B},i_{Q^2},j_z,j_{p_{T}})} \right\vert_{DIS_{rec}}
\label{accp}
\end{equation}

Only 3 bins in $x_B$ and $Q^2$ and 2 bins in $z$ are yet considered, the limits of these bins are presented in tab. \ref{tab1}.	

\begin{table}[htdp]
\centering
\caption{($x_B$,$Q^2$) bins vs. which $M^{h^{\pm}}$ are extracted.}
\label{tab1}
\begin{tabular}{c|p{1.6cm}cp{0.75cm}p{0.75cm}}
\hline \hline
bin n. &  [$x_{B,min}$, $x_{B,max}$] & [$Q^2_{min}$, $Q^2_{max}$] & <$Q^2$> & <$x_B$> \\ \hline
1	 & $0.006,0.008$ & $[1.3,1.7]$	& 1.48 	& 0.007\\
2	 & $0.018,0.025$ & $[1.5,2.5]$	& 1.92	& 0.021\\
3	 & $0.040,0.070$ & $[3.5,6.0]$	& 4.57	& 0.054\\ \hline\hline
	 & [$z_{min}$,$z_{max}$] & <$z$> & & \\ \hline
1	 & 0.2,0.3	& 0.25 & & \\
2	 & 0.3, 0.6	& 0.45 & & \\ \hline \hline
\end{tabular}
\end{table}

The overall acceptance correction factor is $\le 0.35$.  In addition to the acceptance effects, the measured multiplicities are corrected for radiative effects (only $x_B$ and $y$ dependent) which introduce a bias to the kinematics of affected DIS events, mainly at large $y$ ($y>0.7$ and small $x_B$ ($x_B < 0.01$) regions.  

The resulting multiplicities are shown in fig. \ref{MulFitp} (\ref{MulFitm}) for positive (negative) hadrons, in three different kinematic bins (tab.\ref{tab1}) and in two $z$ bins. Experimental data points are fitted with 1-exponential and 2-exponentials functions versus $p_{T}^2$, showing that the shape of experimental data can not be described with a Gaussian parametrisation in the perspective of extracting TMD PDFs and FFs. \\

\section{Hadron Azimuthal Asymmetries}
\label{sec-3}

With the inclusion of the kinematic dependence on the hadron azimuthal angle $\phi_h$, which is the angle between the lepton scattering plane and the plane defined by the hadron and the virtual photon directions, the spin-averaged cross section in the one-photon approximation \cite{Bacchetta:2006tn} is given by:

\begin{figure*}[htdp]
\centering 
\subfigure{\includegraphics[height=4.3cm,width=.75\textwidth]{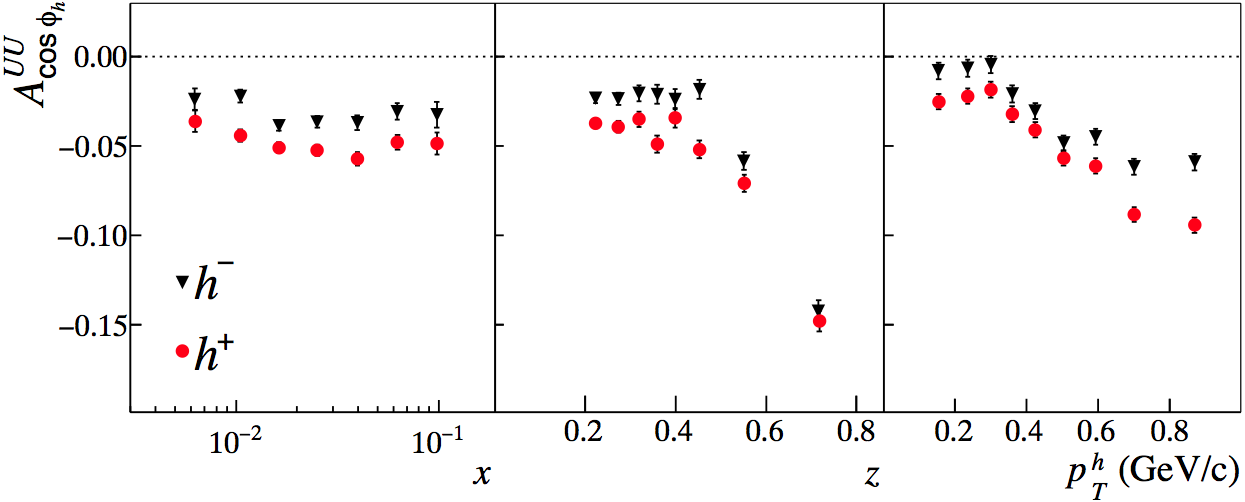}}\vspace{-1.1cm}
\subfigure{\includegraphics[height=4.3cm,width=.75\textwidth]{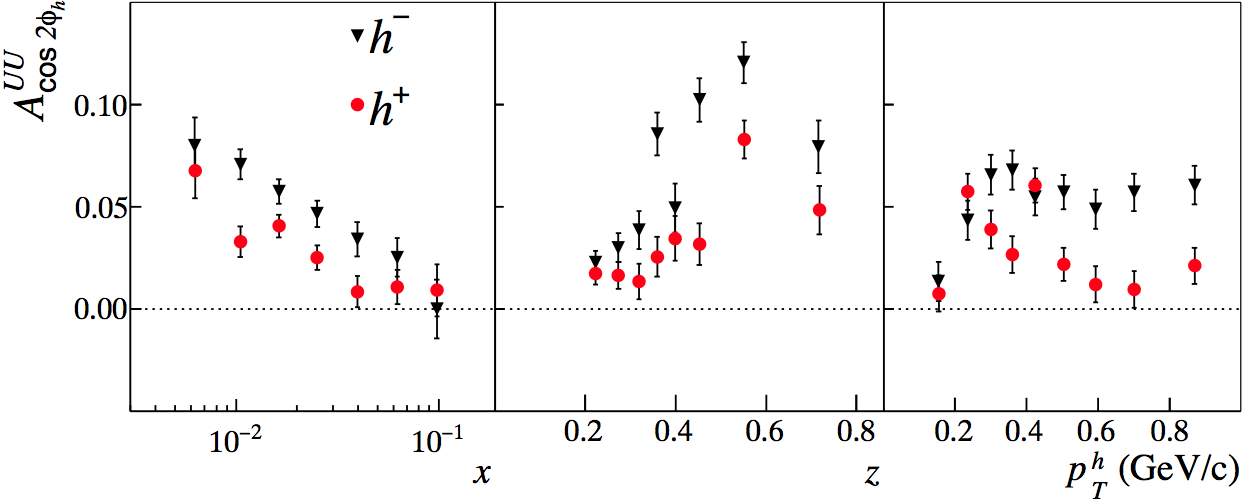}}\vspace{-1.1cm}
\subfigure{\includegraphics[height=4.3cm,width=.75\textwidth]{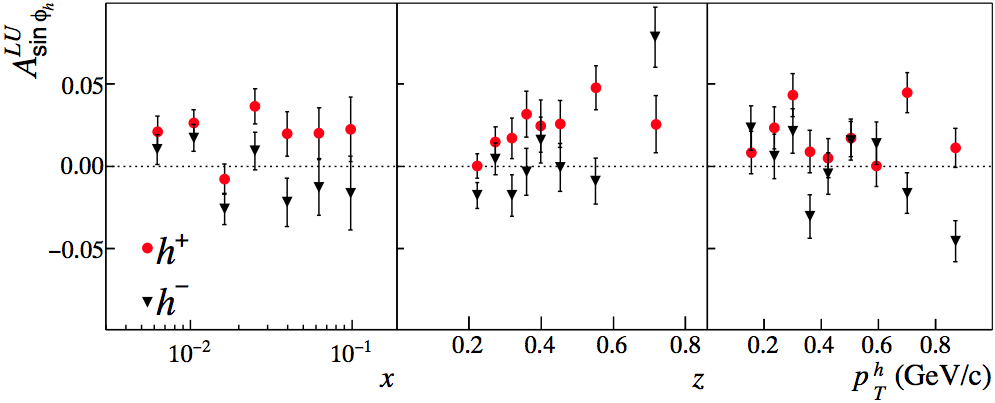}}
\caption{$\cos^{UU}_{\varphi_h}$ and $\cos^{UU}_{2\varphi_h}$ and $\sin^{LU}_{\varphi_h}$ azimuthal  asymmetries for charged unidentified hadrons ($h^{\pm}$).} \vspace{-0.2cm}
\label{cos12phi}
\end{figure*}

\begin{equation}
\begin{gathered}
\label{azi}
\frac{d\sigma}{p_Tdp_Tdx_Bdydzd\phi_h} = \sigma_0 (1 + \epsilon_1 A^{UU}_{\cos\phi_h} \cos\phi_h + \\ 
\epsilon_2 A^{UU}_{\cos 2\phi_h} \cos2\phi_h + \lambda\epsilon_3 A^{LU}_{\sin\phi_h}\sin\phi_h)
\end{gathered}
\end{equation}

where $\alpha$ is the fine structure constant, $\sigma_0$ is the $\phi_h$ independent part of the cross section, $\lambda$ is the longitudinal polarisation of the incident lepton and the quantities $\epsilon_i$ are depolarisation factors dependent upon $y$, $\epsilon_1=(2(2-y)\sqrt(1-y))/(1+(1+y)^2)$ and $\epsilon_2 = (2(1-y))(1+(1-y)^2)$. As seen in eq.\ref{azi}, the cross section exhibit a $\cos\phi_h$ and a $\cos2\phi_h$ azimuthal modulations. They are generated by the so called Cahn effect  \cite{Cahn:1978se}, which arises from the fact that the kinematics is non collinear when the $k_{\perp}$ is taken into account (i.e. a kinematical higher twist), and with the perturbative gluon radiation, resulting in order $\alpha_{s}$ QCD processes. pQCD effects are becoming important for high transverse momenta of the produced hadrons, while they are small for $p_T$ up to 1 $GeV/c$. For the extraction of unpolarised asymmetries the usual tricks to cancel the experimental acceptance are not applicable and a full acceptance correction is needed in order to extract the amplitude of the azimuthal modulations.
On top of the Cahn effect come the particularly interesting so called Boer-Mulders function, describing the transverse parton polarisation inside an unpolarised hadron and generating azimuthal asymmetries in this part of the SIDIS cross section. It is a T-odd function and describes the correlation between the quark transverse spin and its transverse momentum in an unpolarised nucleon \cite{Boer:1997nt}. The Boer-Mulders PDFs contribute to both the $cos\phi_h$ and the $\cos2\phi_h$ modulations. 

Unpolarized azimuthal asymmetries have been measured by the EMC collaboration \cite{EMC1,EMC2}, with a liquid hydrogen target and a muon beam at a slightly higher energy, but without charge separation for the final state hadrons, an important aspect since the Boer-Mulders contribution is expected to have opposite sign or opposite charges. These data have been used \cite{Ansel} to extract the average $<k_{\perp}^2>$. Azimuthal asymmetries have been also measured by E665 \cite{E665} and at higher energies by ZEUS \cite{Zeus}. More recent are the measurements done by HERMES \cite{hermes} and by COMPASS \cite{Adolph:2014pwc}, presented here.

To extract the asymmetries, the azimuthal hadron distributions $N\left( \varphi_h \right)$ are reproduced, for each hadron charge, in each kinematic bin and in 16 bins in $\varphi_h$ for the range $\left[0,2\pi\right]$. These distributions are then corrected for the apparatus acceptance, estimated using a Monte Carlo simulation, and fitted with a four parameters function. More details about the extraction method and systematic studies can be found in \cite{Adolph:2014pwc}.

\begin{figure*}[htdp]
\centering
\includegraphics[height=15cm,width=.8\textwidth]{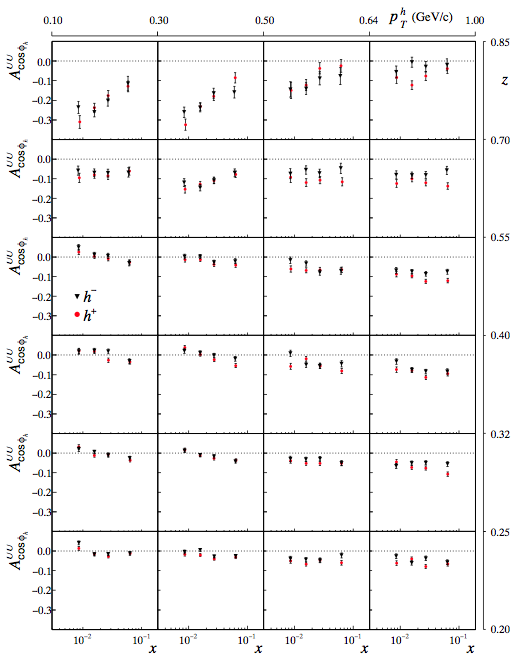}
\caption{$A_{UU}^{\cos\varphi_h}$ asymmetries for $h^{\pm}$ as a function of $x_B$ for different bins in $z$ (from bottom to top) and $p_T^h$ (from left to right). The error bars correspond to statistical uncertainties.}
\label{cosphi3d}
\end{figure*}
\begin{figure*}[htdp]
\centering
\includegraphics[height=15cm,width=.8\textwidth]{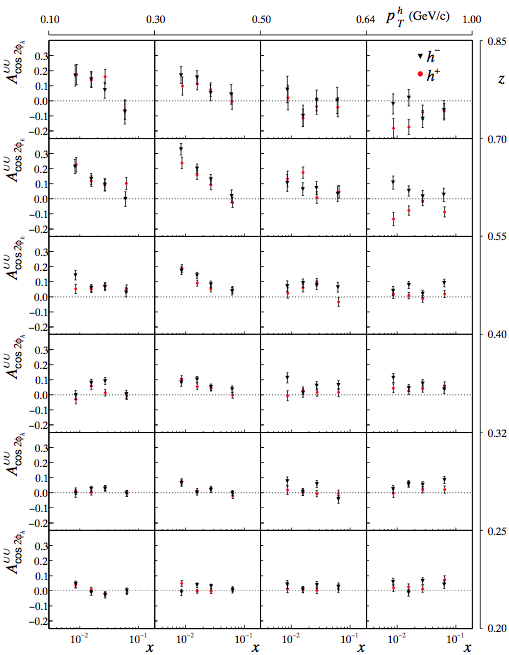}
\caption{ $A_{UU}^{\cos2\varphi_h}$ asymmetries for $h^{\pm}$ as a function of $x_B$ for different bins in $z$ and $p_T^h$, same as fig. \ref{cosphi3d}.}
\label{cos2phi3d}
\end{figure*}

The $\cos\phi_h$ ($A^{UU}_{\cos\phi_h}$), $\cos2\phi_h$ ($A^{UU}_{\cos2\phi_h}$) and $\sin\varphi_h$ ($A^{LU}_{\sin\varphi_h}$) asymmetries are shown in fig.\ref{cos12phi} for both hadron charges versus $x_B$, $z$ and $p_T^h$. The error bars represent statistical uncertainties and the systematic uncertainties are estimated to be as large as twice the statistical ones \cite{Adolph:2014pwc}. 
The $A^{LU}_{\sin\varphi_h}$ is found to be small and compatible with zero for $h^-$. For $h^+$ however, it is slightly positive and increasing with $z$.  
The $\cos\varphi_h$ and $\cos2\varphi_h$ asymmetries are found to be significantly different from zero. The $A^{UU}_{\cos\phi_h}$ asymmetry is negative for both hadron charges with a larger absolute values for positive hadrons and strongly depends on kinematic variables, in particular $z$ and $p_T^h$. Versus $z$, it has a constant value u to $z \eqsim 0.5$ and increase at larger $z$ up to $0.15$ in absolute value. Versus $p_T^h$ it shows a similar behaviour being constant up to $p_T^h \simeq 0.35$ GeV/c and then rapidly increases in absolute value. 
The $A^{UU}_{\cos2\phi_h}$ is found to be significantly positive for both charges with a larger absolute values for negative hadrons. It also shows a strong kinematic dependence. While it shows a descending trend versus $x_B$, it decreases versus $z$ and $p_T^h$ up to $z \simeq 0.6$ and $p_T^h \simeq  0.4$ GeV/c. 
First theoretical attempts to describe the observed strong kinematic dependencies in terms of Cahn and Boer-Mulders effects could not reproduce the trends suggested by experimental results in particular the $p_T^h$ dependence. 

\begin{table}[htdp]
\centering
\caption{$z$, $p_T$ and $x_B$ bins used in the azimuthal asymmetries.}
\label{tab2}
\begin{tabular}{ccc}
\hline \hline
 z & $p_T$ & $x_B$ \\ \hline
0.20-0.25	& 0.10-0.30	& 0.003-0.012\\ 
0.25-0.32	& 0.30-0.50	& 0.012- 0.02\\
0.32-0.40	& 0.50-0.64	& 0.02-0.038\\
0.40-0.55	& 0.64-1.00	& 0.038-0.13\\
0.55-0.70	& & \\
0.70-0.85	& & \\ \hline \hline
\end{tabular}
\end{table}

A deeper investigation of the observed kinematic dependencies has been performed by extracting the azimuthal modulations binning simultaneously the data in $x_B$, $z$ and $p_T^h$ (3-dimensional binning as shown in tab. \ref{tab2}). The results for $A^{UU}_{\cos\varphi_h}$ and $A^{UU}_{\cos2\varphi_h}$ are shown, versus $x_B$ in bins of $z$ and $p_T^h$, in Figs. \ref{cosphi3d} and \ref{cos2phi3d}  respectively, where the error bars represents statistical uncertainties. It has been checked that projecting the asymmetries on any of the three kinematic variables reproduce the results for the integrated asymmetries, as expected. For $A_{\sin\varphi_h}^{LU}$, no specific effect could be observed due to large statistical errors and the results are not shown here.
The large negative signal of $A^{UU}_{\cos\phi_h}$ at small x is originated by high $z$  and small $p_T^h$ hadrons, i.e. $0.55 < z < 0.8$ and 0.1 GeV/c $< p_T^h < 0.64$ GeV/c. In addition, the absolute values of $A^{UU}_{\cos\varphi}$ increases with $p_T^h$ for $z < 0.55$. 
For $A^{UU}_{\cos2\varphi_h}$ asymmetry, its large positively increasing signal versus $x_B$ is originated by high $z$ and small $p_T^h$ hadrons, i,e, $0.55 < z < 0.85$ and 0.1 GeV/c $< p_T^h < 0.5$ GeV/c. For $p_T^h > 0.5$ GeV/c, the asymmetry shows a linear $x_B$ dependence. The difference between positive and negative hadron charges is mainly seen at large $p_T^h$ (0.64 GeV/c $< p_T^h <1$ GeV/c). 
To summarise, the experimental data shows different trends in the various regions in ($z$,$p_T^h$) suggesting that there are different regimes with different dominant processes. Thus, a new phenomenological approach taking into account these observations is needed for a better description of the experimental data and a deeper understanding of the physics behind.

\section{Conclusions}
\label{sec-3} 

COMPASS has produced new inputs for global QCD analysis to extract the intrinsic transverse momenta of quarks. The first are the $p_T^2$-dependent hadron multiplicities measured using data collected in 2006 by scattering a 160 GeV muon beam off a deuteron ($^6$Lid) target. Yet, only part of the total set is shown in three ($x_B$,$Q^2$) kinematic bins and 2 $z$ ranges. However the total set covers a wide kinematic domain in $x_B$ down to 0.003 and $Q^2$ up to 30 (GeV/c)$^2$ and covers a $p_T^2$ range up to 3 (GeV/c)$^2$. The experimental data clearly need 2-exponentials function to produce their shape.
The second input are the azimuthal asymmetries measured using 2004 data. Two sets of results are produced, the first by binning the data in $x_B$,$z$ and $p_T^h$ integrating over the other two variables and the second in a three-dimensional grid of the three variables. The asymmetries of the $\cos\phi_h$ and $\cos2\varphi_h$ modulations show strong kinematic dependencies which could not be reproduced by the existing phenomenological approach. These results constitute an important input to improve the understanding of the transverse momentum structure of the nucleon.

\section{Ackonwledgments}
The author likes to thank the organisers of the conference for the invitation. Finally, this work has been undertaken with the support of the ICTP TRIL programme, Trieste, Italy.



\begin{thebibliography}{}

\bibitem{Adolph:2013stb} 
  C.~Adolph {\it et al.}  [COMPASS Collaboration],
  \href{}{Eur.\ Phys.\ J.\ C {\bf 73}, 2531 (2013)}
  
\bibitem{Airapetian:2012ki} 
  A.~Airapetian {\it et al.}  [HERMES Collaboration],
  \href{}{Phys.\ Rev.\ D {\bf 87}, 074029 (2013)}
  
\bibitem{Anselmino:2013lza} 
  M.~Anselmino, M.~Boglione, J.~O.~Gonzalez H., S.~Melis and A.~Prokudin,
  \href{}{JHEP {\bf 1404}, 005 (2014)}
  
\bibitem{Bacchetta:2006tn} 
  A.~Bacchetta, M.~Diehl, K.~Goeke, A.~Metz, P.~J.~Mulders and M.~Schlegel,
  \href{http://iopscience.iop.org/1126-6708/2007/02/093/}{JHEP {\bf 0702}, 093 (2007)}
  
\bibitem{Cahn:1978se} 
  R.~N.~Cahn,
  \href{http://www.sciencedirect.com/science/article/pii/0370269378900205}{Phys.\ Lett.\ B {\bf 78}, 269 (1978).}
  
\bibitem{Boer:1997nt} 
  D.~Boer and P.~J.~Mulders,
  \href{http://journals.aps.org/prd/abstract/10.1103/PhysRevD.57.5780}{Phys.\ Rev.\ D {\bf 57}, 5780 (1998)}
  
\bibitem{EMC1} Arneodo M \textit{et al.} 1987 Z. Phys. C 34 227
\bibitem{EMC2} Ashman J \textit{et al.} 1991 Z. Phys. C 52 361
\bibitem{Ansel} Anselmino M \textit{et al.}  2005 Phys. Rev. D 71 74006
\bibitem{E665} Adams M \textit{et al.} 1993 Phys. Rev. D 48 5057
\bibitem{Zeus} Bretiwef J \textit{et al.} 2000 Phys. Lett. B 481 199  
\bibitem{hermes} HERMES Collaboration (Giordano F) 2009 AIP Conf. Proc. 1149 423
\bibitem{Adolph:2014pwc} 
  C.~Adolph {\it et al.}  [COMPASS Collaboration],
  \href{http://www.sciencedirect.com/science/article/pii/S0550321314002387}{Nucl.\ Phys.\ B {\bf 886}, 1046 (2014)}

\end{thebibliography}
\end{document}